\documentstyle[aps,twocolumn,psfig]{revtex}
\begin{document}
\draft
\title{Monte Carlo Renormalization of the 3-D Ising model: Analyticity
and Convergence}
\author{H.W.J. Bl\"{o}te, J.R. Heringa, A. Hoogland, E.W. Meyer and T.S. Smit}
\address { Department of Applied Physics, Delft University of Technology, \\
Lorentzweg 1, 2628 CJ Delft, The Netherlands \\ }
\date{\today}
\maketitle
\begin{abstract}

We review the assumptions on which the Monte Carlo renormalization 
technique is based, in particular the analyticity of the block 
spin transformations. On this basis, we select an optimized
Kadanoff blocking rule in combination with the simulation 
of a $d=3$ Ising model with reduced corrections to scaling.
This is achieved by including interactions with second and third 
neighbors. As a consequence of the improved analyticity properties,
this Monte Carlo renormalization method yields a fast convergence 
and a high accuracy. The results for the critical exponents are 
$y_H=2.481(1)$ and $y_T=1.585(3)$.

\end{abstract}
\pacs{02.70.Lq, 05.50.+q, 75.10.Hk, 75.40.Mg}

Applications of the Monte Carlo renormalization group (MCRG) to the
three-dimensional Ising model~\cite{BS79,EMCRG,DMCRG,BGHP}
have become increasingly elaborate and complicated, and tend to 
require considerable computer resources. Nevertheless, there
are still uncertainties due to the basic assumptions underlying
the renormalization transformations used. In particular we will
focus on the question concerning the analyticity of the transformation,
which is closely related to the question whether the corrections to
scaling vanish at the fixed point of the transformation.\cite{Fisher}
In this letter we first calculate the analytic part of a divergent
observable; this demonstrates that gross 
nonanalyticities are normally absent. However, in general we may expect
weak nonanalyticities due to corrections to scaling. Thus,
second, we minimize their effects by adjusting the transformation 
as well as the Hamiltonian that is simulated.

The MCRG method has amply been reviewed,\cite{Sw82,gupta} and here we
only briefly outline the method. 
The reduced Hamiltonian is written as
\begin{equation}
{\cal H}(K_0,K_1,K_2,\cdots;S)=-\sum_{\alpha=0}^{\infty} K_\alpha S_\alpha
\label{generalH}
\end{equation}
where $S$ is a spin configuration, the $K_\alpha$ are couplings,
and the $S_\alpha$ are the conjugate lattice sums over spin products, 
{\it e.g.} $K_1$ is the magnetic field and $S_1=\sum_{i} s_i$ the sum 
over all spins;  $K_2$ is the nearest-neighbor coupling and
$S_2=\sum_{<nn>} s_i s_j$ the sum over all nearest-neighbor pairs 
$(s_i,s_j)$. A special 'coupling' is the background 
energy density $K_0$; $S_0$ is the number of spins.  Application of a 
block-spin transformation to Monte Carlo generated configurations $S$ 
leads to configurations $S'$ described by a Hamiltonian 
${\cal H}'={\cal H}(K_0',K_1',K_2',\cdots;S')$. The renormalized 
couplings $K_{\alpha}'$ are {\em assumed} to be analytic functions of 
the original ones, even at the infinite system critical point. 
However, this property remains unproven in general, even for
$K_0'$, or the so-called 'analytic part' of the transformation!
 
It is straightforward to calculate, using Monte Carlo,
\begin{equation}
B_{\alpha \beta}^{(i)} = 
           \langle \langle S_{\alpha}'  S_{\beta}' \rangle \rangle
  \equiv \langle ( S_{\alpha}'- \langle S_{\alpha}' \rangle) 
                 ( S_{\beta}' - \langle S_{\beta }' \rangle) \rangle
\rangle
\label{defB}
\end{equation}
and
\begin{equation}
C_{\alpha \beta}^{(i)} = 
         \langle \langle S_{\alpha}'  S_{\beta} \rangle \rangle
  \equiv \langle ( S_{\alpha}'- \langle S_{\alpha}' \rangle) 
                 ( S_{\beta}  - \langle S_{\beta }  \rangle) \rangle
\label{defC}
\end{equation}
These lattice sum correlations are related~\cite{Sw79} to the linearized 
transformation 
\begin{equation}
T_ {\alpha \beta}=
\partial K_{\alpha}' / \partial K_{\beta}
\label{defT}
\end{equation}
via
\begin{equation}
B_{\alpha \gamma} T_{\gamma \beta} =
C_{\alpha \beta}
\label{solT}
\end{equation}
The dummy index summation rule applies to Greek indices. The matrix 
${\bf T}$ is approximated by solving Eq.~(\ref{solT}) after truncation 
to a finite number of 
couplings. Under iteration of the block-spin transformation, the 
$K_{\alpha}$ ($\alpha>0$) are assumed to approach a critical fixed-point, 
where the eigenvalues of ${\bf T}$ determine the critical exponents. 
 
Thus, the MCRG method relies on assumptions of 1) analyticity 
of the transformation; 2) convergence with the dimensionality of
the coupling subspace; and 3) convergence to a critical fixed point.

Concerning the third assumption, numerical work involving several
subsequent transformations~\cite{EMCRG,DMCRG,BGHP} suggests that 
convergence to a fixed point does occur, and is described by an 
irrelevant exponent $y_i$ in the range $-0.8$ to $-1.0$.

In order to investigate the second assumption, 
the number of couplings $n_c$ used in the analyses 
has increased considerably over the years; from 7 in Ref.~\cite{BS79} to
99 in Ref.~\cite{BGHP}. A criterion to distinguish 'important' and 'less
important' couplings was introduced in Ref.~\cite{DMCRG}.
The 'importance index' of an $n$-spin coupling is given by  
$( 2^{n/2} \: \overline{r} \: )^{-1}$, where 
$\overline{r}$ is the average distance between the spins.
This formula accounts for the facts that
couplings tend to become less important when more spins are involved and
when $\overline{r}$ increases.\cite{DLB} An ordering 
according to this index leads to fast convergence~\cite{DMCRG} with 
increasing $n_c$. On this basis we have restricted the present 
calculations to 20 even and 15 odd couplings, and indeed we observe
good convergence with $n_c$ for all results presented here.

Next, we search for nonanalyticities in the 'analytic' part 
of the transformation when applied to an infinite system. First we 
investigate if the analytic parts of the susceptibility and the 
specific heat of the nearest-neighbor model, which are proportional to 
$T_{011}$ and $T_{022}$ respectively, are bounded at criticality. 
We express these quantities in derivatives of $\ln Z$ and apply the
chain rule:
\begin{equation}
\frac{ \partial^2 \ln Z }{ \partial K_{\alpha} \partial K_{\beta}}=
\frac{ \partial }{ \partial K_{\alpha}} T_{\gamma \beta}
\frac{ \partial \ln Z }{ \partial K_{\gamma}'} =
T_{\gamma \alpha \beta} \frac{ \partial \ln Z }{ \partial K_{\gamma}'} +
T_{\gamma \beta}
\frac{ \partial^2 \ln Z }{ \partial K_{\alpha} \partial K_{\gamma}'}
\label{dzdkdk}
\end{equation}
where 
$T_{\gamma\alpha\beta}=\partial T_{\gamma\beta}/ \partial K_{\alpha}$.
The derivatives of  $\ln Z$ can trivially be expressed in connected
lattice sum correlations:
\begin{equation}
\langle \langle S_{\alpha} S_{\beta} \rangle \rangle =
T_{\gamma \alpha \beta} \langle S_{\gamma}' \rangle +
T_{\gamma \beta} \langle \langle S_{\alpha} S_{\gamma}' \rangle \rangle 
\label{sssp}
\end{equation}
The $T_{\gamma \alpha \beta}$ are the only unknowns in Eq.~(\ref{sssp}); 
the correlations follow from the simulation, and the $T_{\gamma \beta}$ 
from the standard MCRG analysis. There are not enough equations to solve 
for $T_{0mm}$, but we can calculate the quantity $A_{m m}=L^{-3}T_{\gamma 
m m} \langle S_{\gamma}' \rangle$, in
which the effect of a possible divergence of $T_{0m m}$ vanishes 
only in the case of unlikely cancellations. The factor $L^{-3}$ 
normalizes $S_{\gamma}'$ with respect to the system size $L$.

We have done such calculations using the DISP,\cite{disp,DISP} a 
special-purpose computer for Metropolis 
simulations of Ising models. The transformation is 
defined by the probability $P(s')$ of a block spin $s'$: 
$P(s')=\exp(\omega s' s_b)/2\cosh(\omega s_b)$ where $s_b$ is the sum 
of the spins in a $2^3$ block. It approaches the 
majority rule for large $\omega$. In the limit of small  $\omega$, 
the block spins become independent and the critical singularity
moves to the 'analytic' part.\cite{NiemvL}
Numerical results for $A_{11}$ and $A_{22}$ did not suggest divergences
in the analytic parts of the susceptibility, except where expected: for 
small $\omega$.  Fig. 1 shows the numerical results for $A_{11}$, as a 
function of $\omega$ for $L=8$, 16 and 32. 

A more strict test uses explicit calculation of $T_{0m m}$.
We apply the chain rule to the second differentiation in
$ \partial^3 \ln Z / \partial K_{\alpha} \partial K_{\beta} 
\partial K_{\gamma}'$ and express the derivatives of $\ln Z$ again in 
connected lattice sum correlations:
\begin{equation}
\langle \langle S_{\alpha} S_{\beta} S_{\gamma}' \rangle \rangle =
T_{\delta \alpha \beta} 
\langle \langle S_{\gamma}' S_{\delta}' \rangle \rangle +
T_{\delta \beta} 
\langle \langle S_{\alpha} S_{\gamma}' S_{\delta}' \rangle \rangle 
\label{sspspp}
\end{equation}
Choosing $\alpha=\beta=m$ one can solve the unknowns 
$T_{\delta m m}$ $(\delta>0)$ from the numerical data, and thus 
isolate the term with $T_{0mm}$ in Eq.~\ref{sssp}. 
No signs of divergences are seen in the analytic part of the specific 
heat, except for small $\omega$, as illustrated in Fig. 2.

These results are gratifying but the transformation may still have a 
weaker kind of singularity at the infinite system critical point. This can
indeed be expected if the usual corrections to scaling are present at the 
fixed point of the block-spin transformation. As noted by Fisher and
Randeria,\cite{Fisher} in general corrections to scaling vanish only 
at the fixed point of an analytic transformation. Do they vanish at the 
fixed point of a block-spin transformation? This 
seems doubtful, in particular when the transformation contains 
{\em free parameters which move the fixed point over the critical surface}. 
Since the irrelevant fields are absent at a fixed point, any corrections
should be due to some other mechanism. Weak singularities associated with 
corrections to scaling can enter into the 
renormalized Hamiltonian via a weak nonanalyticity of the transformation.

In order to suppress this problem in MCRG, we propose the following
strategy, which is applicable in a more general context than the Ising
model:
\begin{enumerate}
\item 
The Hamiltonian used to generate the Monte Carlo configurations is chosen
such that the corrections to scaling are small.
\item 
The transformation is chosen such that the fixed point is close
to the Hamiltonian mentioned.
\end{enumerate}
To this purpose we included, in addition to nearest-neighbor couplings
$K_2=K_{nn}$, second and third neighbor couplings $K_{2n}$ and $K_{3n}$ in 
the Monte Carlo simulation, and optimized the ratio between the couplings,
and the block-spin parameter $\omega$.

First we used a Monte Carlo Hamiltonian  with $K_{2n}=0$ and 
$K_{3n}/K_{nn}=0.4$, for which the corrections to scaling are 
small.\cite{mcri,BLH} Then the convergence to the fixed point, as apparent 
from the MCRG results for the eigenvalues of $T_{\alpha\beta}$, 
becomes optimal for $\omega=0.4$. This is 
close to a variational value found by Kadanoff et al.\cite{KHY} The 
difference between the Monte Carlo and the fixed-point Hamiltonian follows, 
in a linear approximation,\cite{Sw82,EMCRG} from the difference 
between $\langle S_{\alpha}\rangle$ and $\langle S_{\alpha}'\rangle$ as
determined from separate simulations of systems with compatible sizes.
This calculation was done in the coupling subspace $(K_{nn},K_{2n},K_{3n})$. 

A second approximation of the the $\omega=0.4$ fixed point was found by
using a Monte Carlo Hamiltonian close to the first approximation.
The fixed point was thus estimated 
$(K_{nn},K_{2n},K_{3n})=(0.1109,0.03308,0.01402)$. A finite-size
scaling analysis of Monte Carlo results~\cite{mcri} was used to
determine the critical point more accurately:
$(K_{nn},K_{2n},K_{3n})=(0.1114448,0.0332520,0.0140925)$, with a relative
accuracy of $2 \times 10^{-5}$. This analysis showed that
the corrections to scaling in the Binder cumulant~\cite{Binder} are 
about 6 times smaller than for the 
nearest neighbor Hamiltonian. 

The bulk of the MCRG calculations took place at the estimated critical
point, using system sizes $L=32$, 16 and 8 and lengths of $10^8$, 
$2 \times 10^7$ and $10^7$ sweeps respectively. The sensitivity
to a variation in $K_{nn}$ was estimated from additional runs at
$K_{nn}=0.1114336$ and $K_{nn}=0.1114560$. 

Further details, including the ordering of the couplings according to 
the importance index are contained in Ref.~\cite{DMCRG}.
Table 1 lists the resulting estimates for the exponents
$y_T$ and $y_H$, as determined from the largest eigenvalues of the
$T_{\alpha \beta}$ matrix. Statistical errors were found by 
dividing the runs in 50 subruns. 
Finite-size and renormalization (approach of the fixed point) effects 
were determined with the procedures described e.g. in Ref.~\cite{EMCRG}.
The convergence of $y_T$ and $y_H$ versus $2^{-n}$ is, after correction 
for the finite-size effect, shown in Figs. 3 and 4. For comparison
we include results from Ref.~\cite{DMCRG}, which used the standard
nearest-neighbor Hamiltonian and the majority rule.
Extrapolation of the data for $L=32$ yields our final estimates 
for the critical exponents: $y_H=2.481(1)$ and $y_T=1.585(3)$. These 
results provide a significant improvement over previous MCRG analyses, 
not only concerning the statistical errors, but also the consistency 
with other results for the Ising universality class which are summarized
e.g. in Ref.~\cite{BLH}. The precise agreement of the renormalization
results with those obtained by finite-size scaling confirms the
validity of hyperscaling.\cite{BakKaw}

Earlier attempts to accelerate the convergence used optimized 
transformations.\cite{swendsenopt,GupCor,hasenfratz,gausterer,bennett,mcrg}
The present work uses this idea combined with a Hamiltonian with suppressed 
corrections to scaling. This alleviates the problem with the
assumption of analyticity, and leads to a much improved convergence 
to the fixed point.
The error due to the uncertainty margin of the exponent describing
the renormalization effect practically vanishes. 
Furthermore, the rapid convergence to the fixed point eliminates the 
necessity of time-consuming simulations of large system sizes.
Further improvements of the MCRG method may be 
possible by the introduction of more adjustable parameters in the 
block-spin transformation, so that its fixed point 
can be moved to a point with even smaller corrections to scaling.

%\acknowledgments
This work is part of the research program of the "Stichting voor 
Fundamenteel Onderzoek der Materie (FOM)", which is financially 
supported by the "Nederlandse Organisatie voor Wetenschappelijk 
Onderzoek (NWO)".

\begin{table}
\caption{Numerical results for the renormalization exponents $y_T$ 
and $y_H$, obtained after $n$ block-spin transformations of a system 
of size $L$.}
\label{tabytyh}
\begin{tabular}{||c|c|c|c|c||}
\hline
exponent& $n$  &  $L=32$       &   $L=16$      &   $L=8$     \\
\hline
$y_T$   &  1   & 1.5885     (3)& 1.5885     (6)& 1.5868   (8)\\
$y_T$   &  2   & 1.5852     (5)& 1.5829     (9)&             \\
$y_T$   &  3   & 1.5829     (9)&               &             \\
\hline
$y_H$   &  1   & 2.48492   (~4)& 2.48500   (~7)& 2.48521 (22)\\
$y_H$   &  2   & 2.48309   (11)& 2.48327   (25)&             \\
$y_H$   &  3   & 2.48219   (27)&               &             \\
\hline
\end{tabular}
\end{table}
\pagebreak   
\begin{figure}
\caption{ The quantity $A_{11}$ defined in the text versus the block-spin
parameter $\omega$ for finite sizes $L=8$ ($\Box$), $L=16$ ($\triangle$)
and $L=32$ ($\circ$). The lines are guides to the eye. Signs of a
divergence with $L$ appear only for small $\omega$.}
\label{fig1}
\psfig{figure=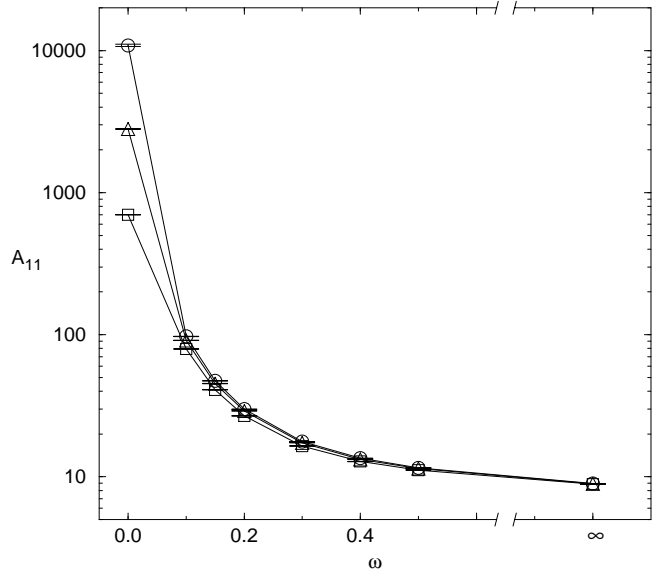,width=\columnwidth}          
\end{figure}
%\pagebreak   
\begin{figure}
\caption{The quantity $T_{022}$, which is proportional to the analytic 
part of the specific heat, versus 
the block-spin parameter $\omega$ for finite sizes $L=8$ ($\Box$), $L=16$ 
($\triangle$) and $L=32$ ($\circ$). The lines are guides to the eye. 
Signs of a divergence with $L$ appear only for small $\omega$.}
\label{fig2}
\psfig{figure=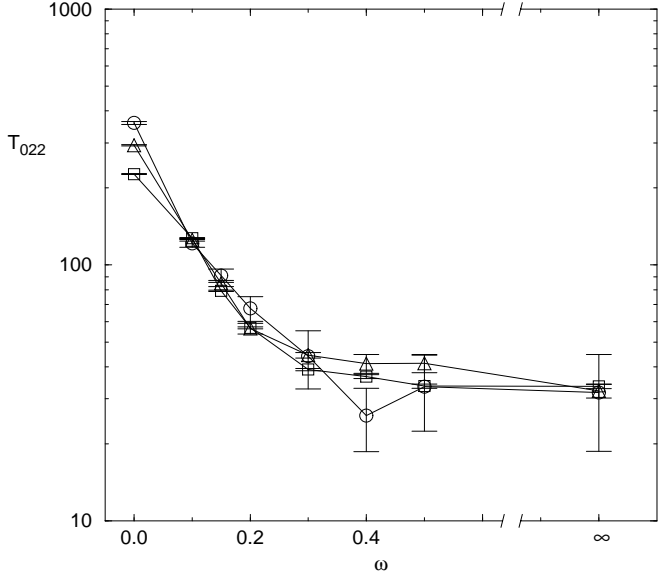,width=\columnwidth}          
\end{figure}
%\pagebreak   
\begin{figure}
\caption{The convergence of the temperature exponent $y_T$ with increasing
number of iterations $n$ of the block-spin transformation. Results are
shown for the present MCRG calculations ($\bullet$) and for those reported
in Ref.~\protect\cite{DMCRG} ($\circ$).}
\label{fig3}
\psfig{figure=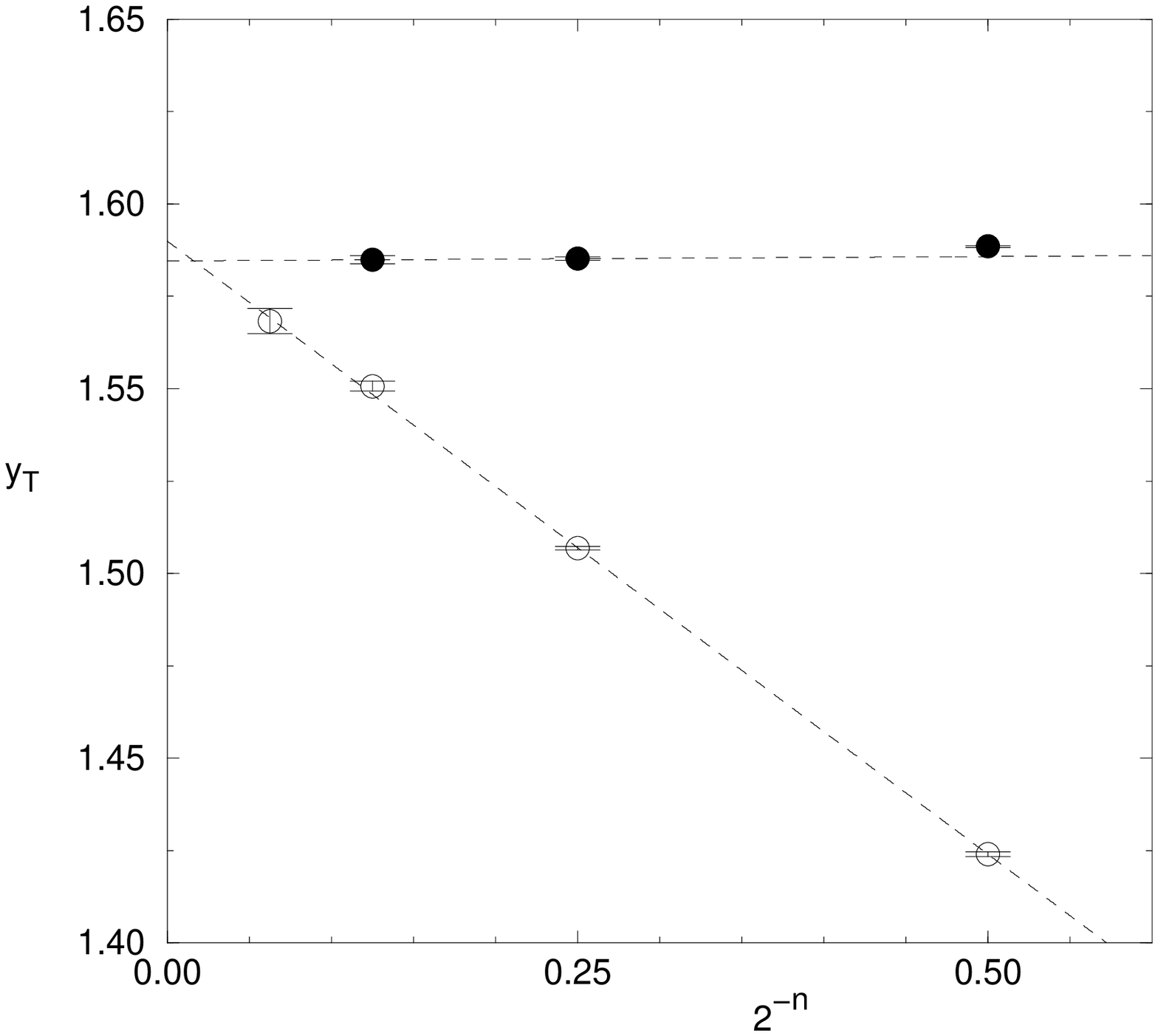,width=\columnwidth}          
\end{figure}
%\pagebreak   
\begin{figure}
\caption{The convergence of the magnetic exponent $y_H$ with increasing
number of iterations $n$ of the block-spin transformation. Results are
shown for the present MCRG calculations ($\bullet$) and for those reported
in Ref.~\protect\cite{DMCRG} ($\circ$).}
\label{fig4}
\psfig{figure=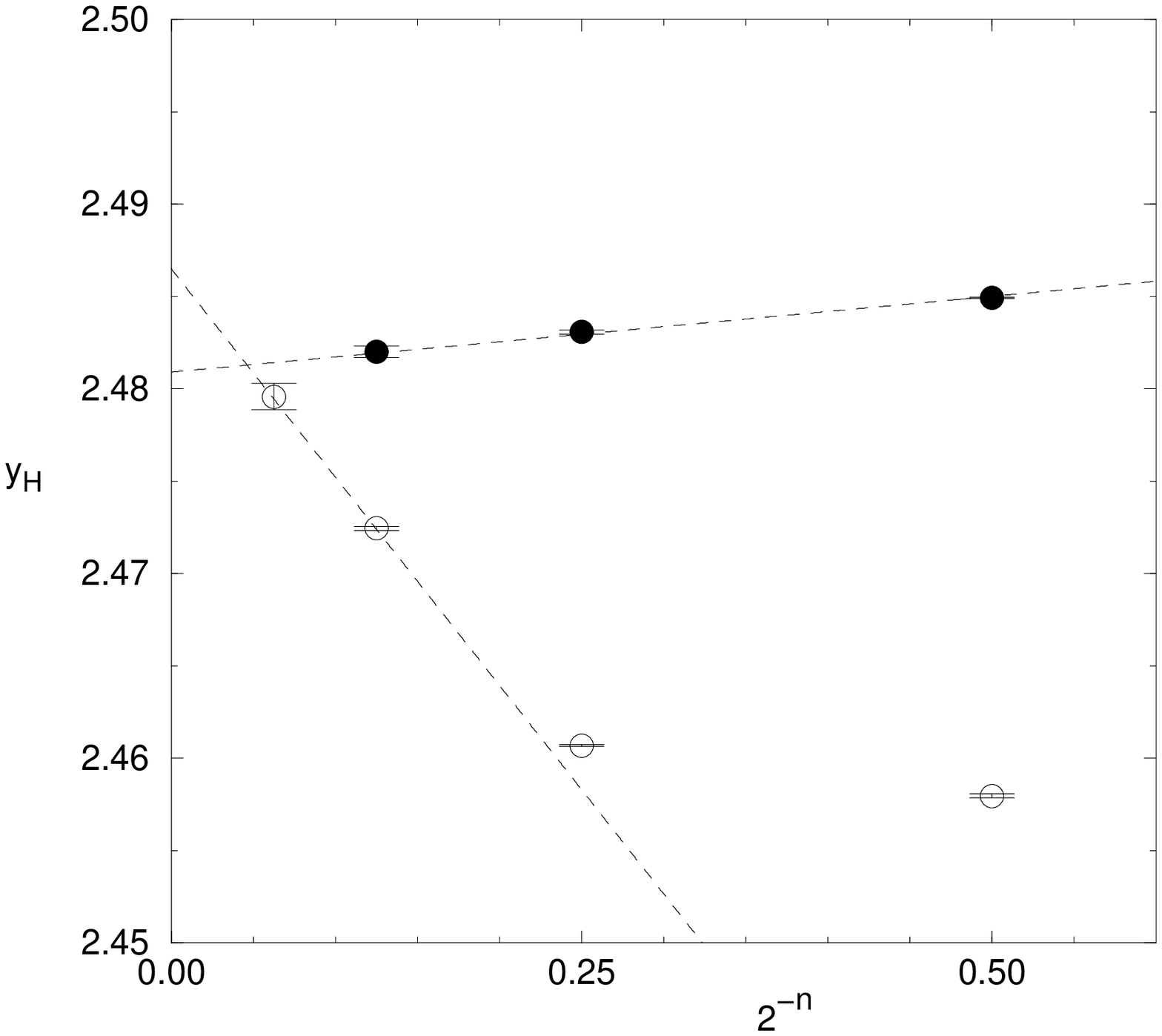,width=\columnwidth}          
\end{figure}
\end{document}